\documentclass[11pt]{article}
\usepackage{epsfig}

\begin{document}

\def\a{\alpha}
\def\b{\beta}
\def\g{\gamma}
\def\G{\Gamma}
\def\d{\delta}
\def\D{\Delta}
\def\e{\epsilon}
\def\h{\hbar}
\def\ve{\varepsilon}
\def\z{\zeta}
\def\t{\theta}
\def\vt{\vartheta}
\def\r{\rho}
\def\vr{\varrho}
\def\k{\kappa}
\def\l{\lambda}
\def\L{\Lambda}
\def\m{\mu}
\def\n{\nu}
\def\o{\omega}
\def\O{\Omega}
\def\s{\sigma}
\def\vs{\varsigma}
\def\S{\Sigma}
\def\vphi{\varphi}
\def\av#1{\langle#1\rangle}
\def\pa{\partial}
\def\na{\nabla}
\def\hg{\hat g}
\def\un{\underline}
\def\ov{\overline}
\def\cF{{\cal F}}
\def\cG{{\cal G}}
\def\cN{{\cal N}}
\def\Hsl{H \hskip-8pt /}
\def\Fsl{F \hskip-6pt /}
\def\cFsl{\cF \hskip-5pt /}
\def\ksl{k \hskip-6pt /}
\def\pasl{\pa \hskip-6pt /}
\def\tr{{\rm tr}}
\def\tcF{{\tilde{{\cal F}_2}}}
\def\tg{{\tilde g}}
\def\shalf{\frac{1}{2}}
\def\nn{\nonumber \\}
\def\w{\wedge}
\def\ra{\rightarrow}
\def\la{\leftarrow}
\def\be{\begin{equation}}
\def\ee{\end{equation}}
\def\ad{\bigtriangledown}
\newcommand{\brr}{\begin{eqnarray}}
\newcommand{\err}{\end{eqnarray}}

\begin{titlepage}

\setcounter{page}{0}

\begin{flushright}
COLO-HEP-42 \\
hep-th/0007nnn \\
February 2001
\end{flushright}

\vspace{5 mm}
\begin{center}
{\large {\bf Remarks on cosmological issues in some string theoretic
brane worlds}}
\vspace{20 mm}

{\large Nikos Irges\footnote{e-mail:
irges@pizero.colorado.edu}}\\
{\em Department of Physics, Box 390,
University of Colorado, Boulder, CO 80309.}\\
\vspace{5 mm}
\end{center}

\vspace{10 mm}

\centerline{{\bf{Abstract}}}
We examine, in the context of certain string compactifications resulting
in five dimensional brane worlds the mechanisms of self/fine tuning of
the cosmological constant and the recovery of standard cosmological
evolution. We show that self tuning can occur 
only as long as supersymmetry is unbroken (unless additional
assumptions are made) and that the
adjustment of the cosmological constant to zero 
after supersymmetry breaking and the recovery of standard
evolution are the same problem verifying 
previously made statements in the context 
of general i.e. not necessarily string
theoretic brane worlds. We emphasize, however, that contrary to general brane
worlds where the above adjustment requires a fine tuning, stringy brane
worlds contain an additional integration constant due to the presence of
the compact space thus allowing the adjustment to be done only with
integration constants.

\end{titlepage}

\newpage
\renewcommand{\thefootnote}{\arabic{footnote}}
\setcounter{footnote}{0}

\section{Introduction}

Recent developments in superstring/M theory and supergravity 
have showed that it is possible to have vacuum states 
such that our world is located on a 3-brane 
embedded in a warped 5 dimensional space-time with some
or all of the extra dimensions compact.  
One such case is the Horava-Witten theory 
\cite{Horava:1996ma} compactified on a compact 
Calabi-Yau (CY) threefold \cite{Lukas:1998tt}. 
The compact space is characterized by a set of
moduli $b_i$ describing its shape (squashing modes) and a modulus $\vphi$, 
specifying its size (the breathing mode). Integrating out the squashing
modes can be done by solving the equations 
\cite{Lukas:1998tt}\cite{Spalinski:2000fz}
(for simplicity we assume that they are constant)
\be \z_{HW} b_i={3\over 2}n_i,\ee
where $\z_{HW}=n_k b^k$ with $n_k$ ($k=1,...,h^{1,1}$) integers. $h^{1,1}$ is
the number of non trivial $1,1$ cycles of the CY. The low energy five
dimensional effective action corresponds to an $S^1/{Z_2}$ orbifold, 
the gauge group carrying three branes being $M5$ branes 
located at the fixed points of the orbifold wrapped around the 
$1,1$ cycles of the CY. 
Consistency of the compactification requires non-trivial four-form field
strength flux which induces the presence of a potential in the five
dimensional bulk action.
There is ${\cal N}=2$ supersymmetry in the bulk and half of it is broken
on the branes. We will call this model the HW brane world model.

Another scenario is to start from IIB string theory, compactify it on a
squashed five sphere \cite{Bremer:1998zp}
\footnote{There is nothing sacred about the squashed five sphere. One could 
presumably replace it with some other Ricci curved space which yields 
the appropriate number of supersymmetries and chirality on the branes and 
thus come up with other examples that have similar properties.} 
(so that only ${\cal N}=2$ supersymmetry is left unbroken) 
and subsequently compactify on an 
$S^1/{Z_2}$ orientifold \cite{deAlwis:2000pr}. Tadpole
cancellation will induce BPS $D3$-branes which can be placed at the fixed
points of the orbifold. The gauge group and the chirality originate
from open string states on the $D3$-branes. 
We can stabilize the squashing modes as before and call the breathing
mode $\vphi$.
Consistency of the compactification
requires non-trivial fluxes which
again induce the presence of a potential in the five dimensional bulk action.
In the following we ignore the squashing of the sphere to avoid
unnecessary complications (however, see Appendix).
The potential in that case depends on two parameters. 
One is $\z_{I_5'}$, the
analogue of $\z_{HW}$ in the HW model and can be computed in terms of the
string scale $M_s$, the radius of the 
sphere $L$ and an integer $n$ as \cite{deAlwis:2000pr}: 
\be \z_{{I_5'}}={n\over {M_s^44\pi L^5}}.\ee 
It is clearly quantized just as $\z_{HW}$ is.
The other parameter is new, 
not present in a Ricci flat compactification such as that of the HW theory,
and it is basically the Ricci scalar $R_5$
of the five sphere. It is not quantized: it can be regarded from the
five dimensional point of view as a continuous integration constant.
We will call this model the ${I_5'}$ brane world model. 
Notice that what we have done here is first compactify on the sphere
and subsequently compactify the resulting five dimensional theory on an
interval accompanied by an orientifold projection. An alternative
route would be to first
orientifold the type IIB theory, which would take us to type I string theory, then
T-dualize to get the type I' theory and finally compactify 
(and wrap the branes and orientifold planes) on 
$CP^2\times S^1$. This
would apparently be precisely the compactification proposed
in \cite{deAlwis:2000pr}, provided that the above steps commute. 
For some more details on this see the Appendix.

There has been a lot of discussion on such Randall and
Sundrum 1 (RS1) \cite{Randall:1999ee} type models. 
\footnote{In ``RS1 type'' in this talk we include models with
scalar fields in the bulk.}
Discussions include the possibility (or not) of
having a zero or nearly zero cosmological constant without fine tuning
\cite{DeWolfe:1999cp}, \cite{Kachru:2000hf}, \cite{deAlwis:2000qc}, 
\cite{Forste:2000ft}, 
(self tuning of the cosmological constant)
and the possibility of recovering a standard cosmological evolution
\cite{Csaki:1999jh}, \cite{Cline:1999ts}, \cite{Binetruy:1999hy},
even though these phenomena do not occur automatically in general RS1
brane worlds: cancellation between the bulk and the brane 
cosmological constants and therefore standard cosmological evolution 
is possible with at least one fine tuning.

In this letter we discuss these issues in the context of the HW
and ${I_5'}$ string models and we hope to clarify first somewhat under what
circumstances the various mechanisms can take place and second, we
argue that string theory provides us with one additional integration
constant compared to non-string theoretic brane worlds
due to the presence of the compact space,
so that the fine tuning can possibly be avoided. 
 
\section{Cosmological issues}

The relevant piece of the five dimensional effective action for these
compactifications is
\be\label{5Daction} S_{bulk} = -{1\over {2\k_5^2}}\int d^5x\sqrt{-G}
\Bigl[{\cal R}-\shalf (\pa\vphi )^2-V(\vphi)\Bigr],\ee
where $\k_5^2={1\over M_5^3}$ with $M_5$ the five dimensional Planck
mass, which is related to the ten or eleven dimensional fundamental
scale, once the details of the compactification are specified.
The potential is 
\be \label{CYpot}V(\vphi)={1\over 6}\z_{HW}^2e^{2\vphi}\ee
for the compactified Horava-Witten (HW) theory and
\be \label{Spot}V(\vphi)=-R_5e^{{4\over 5}{\sqrt{5\over 3}}\vphi}+
8\z_{{I_5'}}^2e^{2{\sqrt{5\over 3}}\vphi}\ee
for the compactified ${I_5'}$ theory\footnote{Strictly speaking, this
potential corresponds to compactification on the round sphere, so it
leaves ${\cal N}=8$ supersymmetry in the bulk and ${\cal N}=4$ on the
branes. By squashing the sphere, we could break half of each of the above
supersymmetries, in which case we would have more terms in the
potential. This simpler form, however, will be sufficient to illustrate
our point.}. The five dimensional space time coordinates are labeled 
by $\{x^{\m},r\}$ and the corresponding metric is $G_{MN}$, with
$M,N=0,1,2,3,5$.  
The action for the branes sitting at $r=0$ and $r=\pi R$, 
in the static gauge, is 
\be \label{braneaction}S_{branes}={1\over {\k_5}^2}\int 
d^4x \sqrt{-g}\Bigl[T_{1}(\vphi )\delta(r-0)+
T_{2}(\vphi )\delta(r-\pi R)\Bigr],\ee 
with $T_i$ the brane tensions and $g_{\m \n}$ the four dimensional
induced metric. 
The above brane action corresponds to an energy momentum tensor
(on the brane at $r=0$ for example, which 
we will take to be our ``Standard Model'' (SM) brane)
$(T_1,T_1,T_1,T_1,0)$, to which we will eventually add some additional
matter in the form $(\r,p,p,p,0)$ for a total energy momentum
tensor on the SM brane  
\be \label{EMtensor}T^{M}_{N}=(-T_1-\r,T_1+p,T_1+p,T_1+p,0).\ee 
The matter density $\r$ and pressure $p$ are assumed to depend only on
time and also to be $\r, p << T_1$. The functional form of the tensions 
before supersymmetry breaking can
be computed directly in the Horava-Witten case to be
\be \label{HWtension}T_1=-T_2=\z_{HW} e^{\vphi}\equiv T(\vphi)\ee 
and in the ${I_5'}$ case it can be fixed by supersymmetry to 
\be \label{Stension}T_1=-T_2=-{\sqrt{5R_5}}
e^{{2\over 5}{\sqrt{5\over 3}}\vphi}+
4\z_{{I_5'}}e^{{\sqrt{5\over 3}}\vphi}\equiv T(\vphi).\ee 
We should note here that up today there does not
exist a string theoretic calculation that justifies the
expression (\ref{Stension}) for the five dimensional effective brane tension.
In fact, by naively pulling back the ten 
dimensional $D3$ brane tension to
five dimensions in the given background, one finds precisely (but only)
the second term in the expression for the five dimensional tension 
(\ref{Stension}). This second term is the one that is analogous to the
(quantized) brane tension term in the HW model, proportional to $\z_{HW}$.
The problem here is that the additional term in the brane tension
proportional to the Ricci scalar can not be reproduced by
straightforward reduction arguments. 
The answer might have to do with effects not
completely understood in orientifolding in a non Ricci flat background. 
\footnote{It might be that a compactification of the HW theory on a
non Ricci flat manifold instead of a CY manifold does not have
this problem. The challenge is to find manifolds that preserve 
${\cal N}=2$ supersymmetry in the 5D bulk. Perhaps some product space of
spheres does this.}
Nevertheless, if the compactification is indeed
supersymmetric as we claim, the effective brane tension has to be as in
(\ref{Stension}). We will assume that this is the case and we leave
the proof for a later work.

After supersymmetry breaking, the tension gets renormalized and we
parameterize the effect of supersymmetry breaking on the $i$'th brane by 
${\not T}= {T(\vphi)+\e_it_i(\vphi)}$  
where $\e_i$ is the scale of supersymmetry breaking and therefore
$\e_i << T$ but also $\r, p << \e_i$. 
In the following, we will assume that the breathing
mode $\vphi$ depends only on the coordinate $r$.
Also, we will denote a time
differentiation by a dot, an $r$ differentiation by a prime and
a differentiation with respect to $\vphi$ by a hat. 
The equations of motion corresponding to the actions (\ref{5Daction}) and 
(\ref{braneaction}) are 
\be \label{Einsteineq} E_{MN}+{1\over 2}G_{MN}
\bigl[{1\over 2}\vphi'^2+V(\vphi)\bigr]-
{1\over 2}\d_{M4}\d_{N4}\vphi'^2+
g_{\mu \nu}T(\vphi)\bigl[\d_0-\d_R\bigr]=0\ee
and 
\be \label{phieq}\vphi''+{(\sqrt{-G})'\over \sqrt{-G}}\vphi'
={\hat V}+2{\hat T}\bigl[\d_0-\d_R\bigr],\ee
where $E_{MN}$ are the components of the Einstein tensor and 
we have defined for simplicity 
$\d (r-0)\equiv \d_0$ and $\d (r-\pi R)\equiv \d_R$. 

First, we make the metric ansatz
\be ds^2=e^{2A(r)}\eta _{\m\n}dx^{\m}dx^{\n}+dr^2,\ee
which corresponds to a warped five dimensional but flat and static four
dimensional world and we set $\r=p=0$.
The Einstein equations naturally split into the part along the 
${\mu \nu}$ directions
and the part along the $r$ direction:
\be \label{Einsteinmunu}E_{\mu \nu}+
{1\over 2}e^{2A}\eta_{\mu \nu} \Bigl[{1\over 2}\vphi'^2
+V+2T(\d_0-\d_R)\Bigr]=0,\ee
\be E_{55}-{1\over 2}\Bigl[{1\over 2}(\vphi')^2-V\Bigr]=0.\ee
The non-vanishing Einstein tensor components are
\be E_{\mu \nu}=e^{2A}\eta_{\mu \nu}(6A'^2+3A'')
\hskip .5in {\rm and} \hskip .5in E_{55}=6A'^2.\ee
The equations of motion, in the bulk, 
after some algebra can be brought in the form
\be \label{eqmotion}A''=-{1\over 6}\vphi'^2, \hskip .3in  
A'^2=-{1\over 12}V(\vphi)+{1\over 24}\vphi'^2, \hskip .3in
\vphi ''+4A'\vphi '= {\hat V}.\ee 
It was suggested in \cite{Skenderis:1999mm} that 
supersymmetry preserving (BPS) solutions 
to the above second order set of equations is the 
(equivalent) first order set
\be \label{firstorder}\vphi'={\hat W}, 
\hskip .3in  A'=-{1\over 6}W, \hskip .3in  
V(\vphi)={1\over 2}{\hat W}^2-{1\over 3}W^2,\ee 
where $W$ is a ``superpotential'' to be computed by solving the last of
the above differential equations. 
We separate these solutions in two classes. One
class will be the special solution that does not contain an integration
constant and we will denote it by $w$. The other will be the general
solution with the integration constant, which we will call ${\not w}$. 
For the HW and the ${I_5'}$ models, we have
respectively,
\be w=\z_{HW} e^{\vphi} \hskip 0.3in {\rm and} \hskip 0.3in w=-{\sqrt{5R_5}}
e^{{2\over 5}{\sqrt{5\over 3}}\vphi}+
4\z_{{I_5'}}e^{{\sqrt{5\over 3}}\vphi}.\ee
Expressions for ${\not w}$ for these two models can be found in
\cite{deAlwis:2000pr} and \cite{deAlwis:2000rt}.
The presence of the branes can be taken into
account by integrating over $r$ equations (\ref{Einsteinmunu}) and the
last of equations (\ref{eqmotion}). Then, one obtains 
for the SM brane the junction conditions 
\be \label{junction}\vphi_0'= {\hat T}_0, \hskip .3in A_0'=-{1\over 6}T_0,\ee
where the subscript $0$ reminds us that these conditions have to be
obeyed, in principle, only at $r=0$ (and at $r=\pi R$, since 
a similar set of junction conditions holds for the other
brane as well). Now, notice the similarity between 
equations (\ref{junction}) and  the first two equations in (\ref{firstorder}).
Clearly, if $T_0=W_0$, the junction
conditions are automatically satisfied. Since,
however, these conditions have to be satisfied 
on both branes simultaneously, a particularly 
convenient way that this can happen is if
the relation $T=W$ holds everywhere between 
and including $r=0$ and $r=\pi R$. In
fact, for the two string theory effective actions we are looking at,
this is exactly the case. For instance, in the HW model we have at the
supersymmetric point $T=w=\z_{HW} e^{\vphi}$ and therefore the junction
conditions (which are just the consistency conditions for a vanishing
cosmological constant on our brane), are satisfied identically without
the use of any integration constants 
\cite{Lukas:1998tt} \cite{deAlwis:2000rt}. 
For the ${I_5'}$ model as well, it
has been shown \cite{Bergshoeff:2000zn} \cite{Cvetic:2000id}
that supersymmetry requires the relation $T=w$
for all $r$. This mechanism is sometimes called
self tuning of the cosmological constant. In these 
string theory compactifications, the self tuning mechanism seems to
be a direct consequence of supersymmetry and therefore not 
very surprising. Unfortunately, when supersymmetry is broken, the
tension $T$ gets renormalized and the 
junction conditions, which still need to
be obeyed, can only be satisfied for some  
superpotential different than $w$, namely ${\not w}$, by the choice of a
set of four integration constants, so that ${\not T}_0={\not w}_0$.
Two independent integration constants come from the solution of 
(\ref{firstorder}), one is the size of the orbifold $R$, and one is
hidden in $\z_{HW}$ ($R_5$) in the HW (${I_5'}$) case.
For now, we will assume that this is
always possible for both the HW and the ${I_5'}$ models.   

Next, we make the slightly more interesting metric assumption
\be ds^2=e^{2A(r)}\bigl[-dt^2+e^{a(t)}d{\bf x}^2\bigl]+dr^2,\ee
which corresponds to a flat three dimensional space whose volume is
changing in time. 
This is not necessarily the most general assumption 
(it is too simple; for more realistic metric assumptions see 
for example \cite{Binetruy:1999hy} and 
references therein) but it is good enough
to demonstrate what we think is the essential point here.
We will also allow for supersymmetry breaking, which
will be taken into account by $T\rightarrow {\not T}$.  
The equations of motion corresponding to 
the $E_{00}$, $E_{ij}$, $E_{55}$ components
of the Einstein tensor, now become
\be \label{infleq1} 3{\dot a}^2-e^{2A}\bigl(
6A'^2+3A''+{1\over 4}\vphi'^2+{1\over 2}V\bigr)
=e^{2A}\bigl({\not T}-\r\bigr)(\d_0-\d_R),\ee
\be \label{infleq2} 3{\dot a}^2+2{\ddot a}^2-e^{2A}\bigl(
6A'^2+3A''+{1\over 4}\vphi'^2+{1\over 2}V\bigr)
=e^{2A}\bigl({\not T}+p\bigr)(\d_0-\d_R),\ee
\be \label{infleq3} 3(2{\dot a}^2+{\ddot a}^2)=e^{2A}\bigl(
6A'^2-{1\over 4}\vphi'^2+{1\over 2}V\bigr),\ee
respectively, and the $\vphi$ equation of motion is still
\be \label{infleq4}\vphi''+4A'\vphi'
={\hat V}+2{\not {\hat T}}\bigl[\d_0-\d_R\bigr].\ee
Equations (\ref{infleq1}) and (\ref{infleq2}) in the bulk immediately
imply that 
\be \label{asol} a(t)=c_1t+c_2\ee
and therefore that they are equivalent in the bulk.
Explicit bulk solutions for $A$ and 
$\vphi$ can be found in \cite{Binetruy:1999hy}
so we will not present any here. 
Instead, we turn to the physics of the SM brane.
Integrating (\ref{infleq1}), (\ref{infleq2}) 
and (\ref{infleq4}) we obtain the junction conditions
\be \label{junctioninfl} A'_0=-{1\over 6}({\not T}_0-\r), 
\hskip 0.3in A'_0=-{1\over 6}({\not T}_0+p),
\hskip 0.3in \vphi'_0={\not {\hat T}_0}.\ee
The $\vphi$ junction condition is unchanged since we assumed that
$\r=\r(t)$. The first two junction conditions are consistent only if 
$p=-\r$, which is the equation of state for vacuum. Consequently, the
equation for the conservation of energy 
\be {\dot \r}=-3H(\r+p)\ee
implies that $\r$ is actually constant in time, i.e. matter does not
flow out of the brane. 
Let us now substitute (\ref{junctioninfl}) into (\ref{infleq3}):
\be {{\dot a}_0^2\over {e^{2A_0}}}=\Bigl({1\over 36}{\not T}_0^2-
{1\over 24}{\not {\hat T}_0}^2+{1\over 12}V\Bigr)-
{1\over 18}\r {\not T}_0,\ee
where we have dropped the term ${\cal O}(\r^2)$. Apparently, standard
cosmological evolution is spoiled by the 
quadratic terms inside of the parenthesis.
If the compactification breaks all the supersymmetry, then the brane
tension does not take the form of its BPS form plus an order $\e$
correction and recovery of standard cosmological evolution can be
achieved by adjusting integration constants to cancel those terms. This
step is basically the RS1 mechanism of canceling the brane against the
bulk cosmological constant to 120 orders of magnitude. 
In our case, however, the compactification  
preserves some supersymmetry. Supersymmetry breaks at a lower scale, 
so we can write ${\not T}_0=T_0+\e_i t_i$.  
Then, by choosing an integration constant so that $e^{2A_0}=1$ and expanding
${\not T}_0$, we obtain, suppressing the brane indices on $\e$ and $t$ 
\be \label{Hubble} H_0^2={1\over 12}\Bigl({1\over 3}T_0^2-
{1\over 2}{\hat T}_0^2
+V\Bigr)+{\e\over 12}\Bigl({2\over 3}T_0t -
{\hat T}_0{\hat t}\Bigr)-\r\Bigl({1\over 18}T_0\Bigr),\ee 
where we have defined the present Hubble parameter $a_0\equiv H_0$,
which by (\ref{asol}) is equal to the constant $c_1$ (i.e. no inflation)
and dropped a term of ${\cal O}(\e \r)$. The three groups of terms 
in the right hand side of (\ref{Hubble}), 
(all the terms inside of the parentheses
being ${\cal O}(1)$) are in decreasing order of
magnitude, since $\r << \e << 1$. On the other hand, the left hand side
$H_0^2$ is basically the observed cosmological constant, which is at
most of the order of magnitude of the last term in
(\ref{Hubble}). Therefore, agreement with the small observed value of the
cosmological constant suggests that the terms of ${\cal O}(1)$ and of 
${\cal O}(\e)$ may cancel separately. 
The cancellation of the ${\cal O}(1)$ terms is
indeed not a problem as we saw in the previous section; $T_0=w$ in both the HW
and the ${I_5'}$ cases and the terms cancel by the original supersymmetry.
For the terms of ${\cal O}(\e)$ to vanish, it must be that
${2\over 3}T_0t\simeq {\hat T}_0{\hat t}$. 
This relation, just as before, can be satisfied generically only via
the adjustment of the available integration constants, 
i.e. we can not escape from the cosmological 
constant problem with supersymmetry, we
can only improve it by some fifty orders of magnitude. Again, this is
not a surprise. Once these terms cancel, we have standard cosmological
evolution with $H_0^2\sim \r$ \cite{Csaki:1999jh} \cite{Cline:1999ts}     
\cite{Binetruy:1999hy}. 
Here is in fact where apparently the ${I_5'}$ model 
seems to have an advantage over the HW
model. Recall that in the ${I_5'}$ model we have four continuous integration
constants \cite{deAlwis:2000pr}, whereas in 
the HW model we have three continuous and one set
of quantized constants (hidden in $\z_{HW}$) 
\cite{deAlwis:2000rt}. It seems very hard to satisfy
the cancellation of the ${\cal O}(\e)$ terms with a set of discrete
parameters. 

We can go one step further if in addition we make
the stronger assumption that the actual functional form of
$t$ is implied by the vanishing of the ${\cal O}(\e)$ term.
This amounts to assuming by analogy to the supersymmetric $T=w$ 
relation that ${\not T}={\not w}$ holds 
identically after supersymmetry breaking.  
Substituting in (\ref{Hubble}) the expressions for
$T_0$ ($=T$) in the HW and ${I_5'}$ models and solving for $t$, we obtain a
constraint on the functional form  
of the supersymmetry breaking correction to the brane tension:
\be t_{HW}(\vphi)\sim e^{{3\over 2}\vphi} \hskip 0.3in {\rm and}\hskip 0.3in
t_{{I_5'}}(\vphi)\sim e^{{7\over 3}\sqrt{3\over 5}\vphi}
\Bigl({4\over {\sqrt{3}}}e^{{2\over 5}\sqrt{5\over 3}\vphi}
+8\sqrt{5\over 3}e^{2\sqrt{5\over 3}\vphi}\Bigr).\ee
It would be interesting to check this last assumption 
using some popular supersymmetry breaking schemes, since 
it would be a realization of a self tuning mechanism even after
supersymmetry breaking\footnote{Even if this is the case though, one has to
worry about later phase transitions that might generate additional
contributions to the cosmological constant}. 

Concluding, in this paper we 
emphasized that in our point of view, the reason that it
is possible to have a vanishing cosmological constant and a standard
cosmological evolution in 
RS1 type string theoretic brane worlds such as the HW and ${I_5'}$ models,
is that the four available integration constants 
are just sufficient to satisfy the four 
junction conditions. The fourth integration constant 
($\z_{HW}$ or $R_5$) that is usually
not present in non-string theoretic brane worlds, is associated with the
existence of the compact space.   

\section{Appendix}

In the Appendix we will give some details of a possible string 
theory realization of the $I_5'$ model. 
We emphasize that even though the supergravity realization of 
this model is rather clear, 
several stages of its string theory realization are still 
at this point at a conjectural level and that is why we put it in an Appendix.
The picture we have in mind is the following:

\begin{center}
\epsfig{file=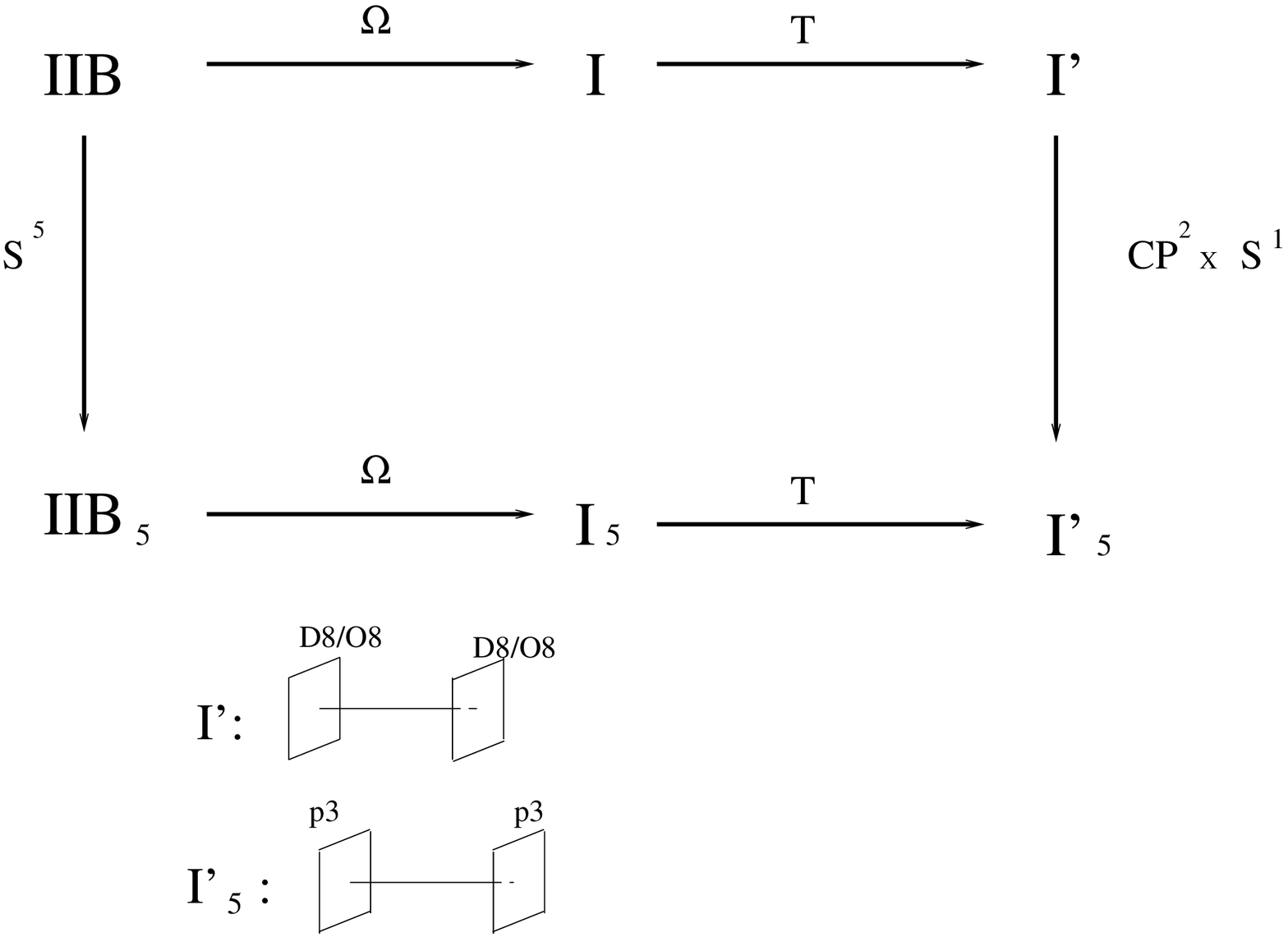,width=12cm}
\end{center}

In the above, we have of course assumed that the diagram is commuting.
Without the intermediate orientifold step $\O$ we know that it is \cite{Duff:1998us}.
With the orientifold step we do not know it but we assume that it is still 
a commuting diagram. The $p3$ branes in the picture at the 
bottmom are the $D8/O8$ branes of the original $I'$ theory wrapped around 
$CP^2\times S^1$. By squashing the $S^5$ on the left, or the $CP^2$ on the right, 
we expect that the maximal supersymmetry will break to 
${\cal N}=2$ in the bulk and to ${\cal N}=1$ on the $p3$ branes in the $I_5'$ theory. 

The the other important issue is whether the $p3$ branes can support 
chiral fermions on their world volume. This is necessary for the $I_5'$ model
to be potentially fully realistic, i.e. to be able to reproduce the 
chiral representations of the quarks and the leptons of the Standard Model.
Let us start again from type $I'$ string theory and compactify two 
directions -$a$ and $b$- along the branes on a torus, 
say $T^2=S^1_a \times S^1_b$. Next, T-dualize along both directions, so that
the $D8/O8$ branes become $D6/O6$ branes. Then compactify two more 
directions along the branes on an $S^2$ and one more direction along the branes,
say direction $c$, on the circle $S^1_c$. Now, the directions $b$ and $c$
form a torus (with one direction along -direction $c$- and the other direction
transverse -direction $b$- to the branes), and the branes can wrap this torus 
in such a way so that they intersect at non-zero angles. 
To summarize, what we have at this point is stacks of $D6/O6$ branes wrapped
around $S^2$, intersecting at angles on a torus, with a transverse 
$S^1_a\times S^1_d/{{Z}_2}$, where the orbifold 
$S^1_d/{{Z}_2}$ is the orbifold of the original $I'$ theory.   
The question here is if this system can support chiral fermions.
Even if it does not, we expect that some simple modification of the model will.
The reason for our expectation is the fact that the $I'$ theory 
is related to the Horava-Witten theory by a simple circle 
compactification along the $E_8$ walls, and we know that the 
HW theory supports chiral fermions on the walls even after 
its compactification on a smooth manifold (eg a Calabi-Yau).
Therefore, we expect that some smooth manifold compactification
of type $I'$ must exist that supports chiral fermions on the branes. 
 
To make the connection with the $I_5'$ model, now T-dualize back along 
directions $a$ and $b$. What we obtain is $D8/O8$ branes wrapped around
$S^2\times T^2 \times S^1_a$, with flux through the torus. 
Furthermore, the $S^2\times T^2$ part need not be globally a trivial product.
It can be a non-trivial bundle for example. If $CP^2$ can be constructed as a torus 
bundle over a 2-sphere (this is not clear at all), then we get precisely to the $I_5'$ model
and if not, to some cousin of it. Notice finally that in the limit 
that the flux goes to zero, which in the T-dual language means that
the branes become parallel, maximal superymmetry is restored.
In other words, tilting the branes in angles must correspond 
to the 'squashing' of the $CP^2$ or of its cousin.     

\vskip 1cm
{\bf Acknowledgements}
\vskip .5cm

I would like to thank Shanta de Alwis for discussions and suggestions.
This work is partially supported by the Department of
Energy contract No. DE-FG02-91-ER-40672.

\bibliography{nikos}

\bibliographystyle{h-elsevier}

\end{document}